\documentclass[12pt]{article}
\usepackage{amssymb}
\begin{document}

\makeatletter
\def\@maketitle{\newpage
 \null
 {\normalsize \tt \begin{flushright} 
  \begin{tabular}[t]{l} \@date  
  \end{tabular}
 \end{flushright}}
 \begin{center} 
 \vskip 2em
 {\LARGE \@title \par} \vskip 1.5em {\large \lineskip .5em
 \begin{tabular}[t]{c}\@author 
 \end{tabular}\par} 
 \end{center}
 \par
 \vskip 1.5em} 
\makeatother
\topmargin=-1cm
\oddsidemargin=1.5cm
\evensidemargin=-.0cm
\textwidth=15.5cm
\textheight=22cm
\setlength{\baselineskip}{16pt}
\title{Fuzzy Torus and q-Deformed Lie Algebra}
\author{  Ryuichi~{\sc Nakayama}\thanks{ nakayama@particle.sci.hokudai.ac.jp}
       \\[1cm]
{\small
    Division of Physics, Faculty of Science,} \\
{\small
           Hokkaido University, Sapporo 060-0810, Japan}
}
\date{
EPHOU-06-001  \\
hep-th/0604010 \\ 
April  2006  
}
%
%
\maketitle

\begin{abstract} 
It will be shown that the defining relations for fuzzy torus and deformed
(squashed) sphere proposed by J. Arnlind, et al (hep-th/0602290) (ABHHS) 
can be rewriten as a new algebra which contains q-deformed commutators. 
The quantum parameter $q$ ($|q|=1$) is a function of $\hbar$. It is shown 
that the $q \rightarrow 1$ limit of the algebra with the parameter $\mu <0$ 
describes fuzzy S$^2$ and that the squashed S$^2$ with $q \neq 1$ and 
$\mu <0$ can be regarded as a new kind of quantum S$^2$. 
Throughout the paper the value of the invariant of the 
algebra, which defines the constraint for the surfaces, is not restricted 
to be 1. This allows the parameter $q$ to be treated as independent of N 
(the dimension of the representation) 
and $\mu$. It was shown by ABHHS that there are two types of representations 
for the algebra, ``string solution'' and ``loop solution''.  
The ``loop solution'' exists only for $q$ a root of unity ($q^N=1$) and 
contains undetermined parameters. The 'string solution' 
exists for generic values of $q$ ($q^N \neq 1$). In this paper we will 
explicitly construct the representation of the q-deformed algebra for 
generic values of $q$ ($q^N \neq 1$) and it is shown that the allowed 
range of the value of $q+q^{-1}$ must be restricted for each fixed N. 

\end{abstract}
\newpage
\setlength{\baselineskip}{18pt}

\newcommand {\beq}{\begin{equation}}
\newcommand {\eeq}{\end{equation}}
\newcommand {\beqa}{\begin{eqnarray}}
\newcommand {\eeqa} {\end{eqnarray}}
\newcommand{\bm}[1]{\mbox{\boldmath $#1$}}
\newcommand{\al}{2\pi \alpha'}

Among fuzzy compact manifolds fuzzy sphere (S$^2$) is the simplest and well-studied 
one.\cite{Madore}\cite{JH}\cite{gaugeaction}
The representation of fuzzy S$^2$ is classified according to the unitary 
irreducible representation of SU(2) Lie algebra and the number of independent 
functions on fuzzy S$^2$ is finite. 
The functions in the spin $j$ representation of fuzzy S$^2$ are mapped onto 
(2j+1) $\times$ (2j+1) matrices and the number of independent functions is 
$(2j+1)^2$. The non-commutative product (star product)\cite{Presnejder}
\cite{HNT}  of functions is one to one correspondence with the product of the 
corresponding matrices.  This property is also true for other fuzzy compact 
manifolds. The representation of fuzzy $\mathbb{C}$P$^2$ corresponds to a 
series of unitary irreducible representations of SU(3) and so forth.\cite{CP2} 

A torus is a compact manifold, but there has been no satisfactory formalism 
for fuzzy torus (T$^{2n}$) in which one can define a star product such 
that the algebra of only a finite number of independent functions is 
closed.\cite{FT}  

Recently, in an interesting paper \cite{FRS} 
a prescription for constructing polynomial relations among 
non-commutative coordinates of compact fuzzy Riemann surfaces 
was proposed and the case of fuzzy T$^2$ was worked out explicitly.
A fuzzy T$^2$ algebra was defined by the eqs.   
\begin{eqnarray}
\ [ X, Y] &=& i \hbar Z, \label{FT1} \\
\ [ Y, Z] & = & i\hbar \left\{X, \ X^2+Y^2 -\mu \right\}, \label{FT2} \\
\ [ Z,X] &=& i\hbar \left\{Y, \ X^2+Y^2 -\mu \right\}, \label{FT3}\\
\ \left( X^2+Y^2-\mu \right)^2+ Z^2 &=& \bm{1}
\label{FT4}
\end{eqnarray}
Here $X$, $Y$, $Z$ are hermitian matrices representing non-commutative 
coordinates and $[A,B]=AB-BA$, $\{A,B \}=AB+BA$. 
$\hbar$, $\mu$ are real parameters.  
For $\mu > 1$ this algebra describes a fuzzy T$^2$ and
for $ -1 < \mu < 1$  a deformed (squashed) fuzzy S$^2$. 
Surprisingly, although the relations (\ref{FT1})-(\ref{FT4}) are complicated,
these satisfy Jacobi identities, finite dimensional representations exist 
and were constructed explicitly in \cite{FRS}.

It was shown that 
\begin{equation}
C= (X^2+Y^2-\mu)^2+Z^2
\label{Casimir}
\end{equation}
commutes with $X$, $Y$, $Z$.\cite{FRS} So this is a multiple of an identity. 
This value does not necessarily be equal to 1 as in (\ref{FT4}).  
Therefore we will set $C=c\bm{1}$ in this paper and let $c$ take an 
arbitrary positive value. Then the relations (\ref{FT1})-(\ref{FT3}), 
(\ref{Casimir}) describe fuzzy T$^2$ for $\mu > \sqrt{c}$ and fuzzy squashed 
S$^2$ for $-\sqrt{c} < \mu < \sqrt{c}$.  Actually the value of $c$ will be 
determined later in terms of $\hbar$, $\mu$ and the dimension N of the 
representation. 

In \cite{FRS} the parameter $\hbar$ was introduced as a quantization parameter 
for replacing Poisson brackets by commutators. In this paper it will be 
regarded as a deformation parameter for deforming a Lie algebra into a 
q-deformed Lie algebra. It will be shown in this paper that when $\mu <0$, the 
$\hbar \rightarrow 0$ limit of the squashed S$^2$ is the ordinary 
fuzzy S$^2$.  

The relations (\ref{FT1})-(\ref{FT3}), are not a Lie algebra and the 
right-hand sides are complicated. In the case of usual fuzzy S$^2$ it is 
natural to embed the sphere in a flat $\mathbb{R}^3$ in an SO(3) invariant way 
because of the SO(3) (SU(2)) symmetry of the fuzzy S$^2$ algebra and the 
constraint $X^2+Y^2+Z^2=R^2 \bm{1}$. In the case of (\ref{FT1})-(\ref{FT4}) 
without symmetry principles except for rotations in $X-Y$ plane it is not 
clear how the T$^2$ and squashed S$^2$ are embedded in 3-dim space. 
The construction of field theories on these fuzzy manifolds is not 
straightforward and there will be ambiguities, even if 
one knows the solutions representing fuzzy configurations. If there
exists symmetry, then it can be used as a guiding principle for constructing 
field theories on fuzzy manifolds. 

One of the purposes of the present paper is to show that the algebra for fuzzy 
T$^2$ (\ref{FT1})-(\ref{FT3}) can be recombined into that which has the 
structure of a q-deformed Lie algebra.  
We will then derive irreducible N-dim representations of this algebra for generic 
values of $q$ ($q^N \neq 1$)(``string solution''). In this 
analysis $q$ defined by (\ref{q}) below is treated as an independent 
q-deformation parameter. According to the sign of $\mu$ the range for the 
value of $q+q^{-1}$ is restricted for each value of N. Then the value of $c$
(\ref{Casimir}) will be determined. The condition on $q+q^{-1}$ for describing 
fuzzy T$^2$ is also obtained. For the representation which exists only for 
$q^N=1$ (``loop solution'') see note added. 

Let us derive the deformed Lie algebra. As in \cite{FRS} we introduce 
$W \equiv X+ iY$ and hermitian matrices$D$, $\tilde{D}$ by 
\begin{equation}
D \equiv W W^{\dagger}, \qquad 
\tilde{D} \equiv W^{\dagger} W.
\label{DD}
\end{equation}
In the complex notation the algebra (\ref{FT1})-(\ref{FT3}) is written as
\begin{eqnarray}
\ [W, W^{\dagger} ] &=& 2 \hbar Z, \label{FTC1}\\
\ [Z, W] &=& \hbar \{ W, \varphi \}, \label{FTC2} \\
\ [Z, W^{\dagger}] &=& - \hbar \{ W^{\dagger}, \varphi \}, \label{FTC3}
\end{eqnarray}
where the matrix $\varphi$ is defined by  
\begin{equation}
\varphi \equiv W^{\dagger}W+\hbar Z-\mu =WW^{\dagger} - \hbar Z-\mu
\label{phi}
\end{equation}
and it commutes with $Z$. \cite{FRS} 
By using (\ref{FTC2}) and (\ref{phi}) one obtains  
\begin{eqnarray}
\ [Z, W] &=& \hbar \left( W(W^{\dagger}W+\hbar Z-\mu)+ 
(WW^{\dagger}-\hbar Z-\mu) W \right) \nonumber \\
&=& 2\hbar (WW^{\dagger}W-\mu W) - \hbar^2 [Z,W] \nonumber \\
&=& \frac{2\hbar}{1+\hbar^2} \ (WW^{\dagger}W-\mu W)
\label{ZW}
\end{eqnarray}  
One can eliminate $Z$ by using (\ref{FTC1}). By expanding the commutators 
one obtains the eq.\cite{FRS} 
\begin{equation}
2\alpha WW^{\dagger}W -W^2W^{\dagger}-W^{\dagger}W^2=
-\frac{4\hbar^2\mu}{1+\hbar^2} \ W
\label{WWW}
\end{equation}
Here $\alpha=(1-\hbar^2)/(1+\hbar^2)$.

Now let us define a new parameter $q$ by 
$q+q^{-1}=2\alpha$. Because $-1 < \alpha \leq 1$, $q$ is a complex number 
\begin{equation}
q=\alpha \pm i \sqrt{1-\alpha^2}= (1-\hbar^2 \pm 2i\hbar)/(1+\hbar^2)
\label{q}
\end{equation}
and its absolute value is $|q|=1$. 
Now we will use (\ref{DD}). 
Because $WW^{\dagger}W$ can be expressed in two ways, $DW$ or $W\tilde{D}$, 
we have from eq (\ref{WWW}) 
\begin{equation}
(qDW + \frac{1}{q}W\tilde{D})-WD-\tilde{D}W= -\frac{4\hbar^2\mu}{1+\hbar^2} \ W
\label{qDW}
\end{equation}
Defining a q-deformed commutator
\begin{equation}
\ [ \ A, \ B \ ]_q \equiv qAB-BA, 
\label{qCom}
\end{equation}
we can put this into the following form. 
\begin{equation}
\ [ \ D-\frac{1}{q}\tilde{D}, \  W \, ]_q= -\frac{4\hbar^2\mu}{1+\hbar^2} \ W
\label{qalgebra1}
\end{equation}
Similarly from eq (\ref{FTC3}) we also obtain the eq. 
\begin{equation}
\frac{1}{q} \ [ \  D- q \tilde{D}, \  W^{\dagger} \, ]_q= \frac{4\hbar^2\mu}{1+\hbar^2} \ W^{\dagger}
\label{qalgebra2}
\end{equation}
Further, we note that if we interchange $q$ and $q^{-1}$ in eq (\ref{qDW}), 
we obtain another set of eqs.
\begin{eqnarray}
\frac{1}{q} \ [ \ W, \ D-q\tilde{D} \, ]_q &=& \frac{4\hbar^2\mu}{1+\hbar^2} \ W, \label{qalgebra4} \\
\ [ \ W^{\dagger}, \ D-\frac{1}{q} \tilde{D} \, ]_q &=& -\frac{4\hbar^2\mu}{1+\hbar^2} \ W^{\dagger}
\label{qalgebra5}
\end{eqnarray} 
We also consider ther relations among $W$, $W^{\dagger}$, $D$, $\tilde{D}$.
\begin{eqnarray}
\ [W, \ W^{\dagger}]_q= qD-\tilde{D}, \label{qalgebra3} \\
\ [W^{\dagger}, \ W]_q= q \tilde{D}-D, \label{qalgebra7} \\
\ [D, \tilde{D}]=0 \label{qalgebra6}
\end{eqnarray}
Note that the commutator in  (\ref{qalgebra6}) is the ordinary one. 
The derivation of these eqs is straightforward due to (\ref{DD}). 
The last eq was proved in \cite{FRS}. 
It is easy to show that the algebra (\ref{FTC1})-(\ref{FTC3}) can be derived 
from  (\ref{qalgebra1})-(\ref{qalgebra6}). If $q \neq \pm 1$, 
which we assume, from (\ref{qalgebra3}) and (\ref{qalgebra7}) one can derive 
$D=WW^{\dagger}$ and $\tilde{D}=W^{\dagger}W$. 
By substituting these eqs into (\ref{qDW}) and using a definition 
$Z =(D-\tilde{D})/2\hbar$,  we obtain (\ref{ZW}). 
Then it is straightforward to derive (\ref{FTC1})-(\ref{FTC3}). 

Now we adopt the eqs (\ref{qalgebra1})-(\ref{qalgebra6}) as the defining 
algebra for the fuzzy T$^2$. This algebra must be supplemented with two 
constraint eqs.  
\begin{equation}
D+\tilde{D}=WW^{\dagger}+W^{\dagger}W
\label{DDWWWW}
\end{equation}
and a quadratic one
\begin{eqnarray}
c \bm{1} 
&=& \frac{1}{4}(D+\tilde{D}-2\mu)^2+\frac{1}{4\hbar^2}(D-\tilde{D})^2 \nonumber \\
&=& \frac{1}{4}(D+\tilde{D})^2-\mu (D+\tilde{D})+\mu^2+\frac{1}{4\hbar^2}(D-\tilde{D})^2 \nonumber \\
&=& -\frac{q}{2(q-1)^2} \ (KK^{\dagger}+K^{\dagger}K)-\mu (WW^{\dagger}+W^{\dagger}W)+\mu^2, 
\label{CMetric}
\end{eqnarray}
which is derived from the invariant (\ref{Casimir}).
Here $c$ is a positive number and $K \equiv D-q\tilde{D}$, 
$K^{\dagger} \equiv D-q^{-1}\tilde{D}$. Eqs $X^2+Y^2=(WW^{\dagger}
+W^{\dagger}W)/2=(D+\tilde{D})/2$ and  $Z=(D-\tilde{D})/2\hbar$ are used to 
transform (\ref{Casimir}) into a quadratic form. Actually, the first 
constraint (\ref{DDWWWW}) can be derived from (\ref{qalgebra3}) and 
(\ref{qalgebra7}) as explained above.

The right-hand sides of (\ref{qalgebra1})-(\ref{qalgebra6}) are at most 
linear in $W$, $W^{\dagger}$, $D$, $\tilde{D}$ and this algebra has a 
structure of a q-deformed Lie algebra. This algebra is similar to the 
q-deformed SU(2) Lie algebra for the quantum 
S$^2$.\cite{qSphere}\cite{qsphereWZ}
Actually, our algebra is a q-analog of SU(2) (or SU(1,1)) Lie algebra 
with \lq doubled' Cartan subalgebra ($D$, $\tilde{D}$). 
There may exist a quantum group symmetry associated 
with this new q-deformed Lie algebra.   

There is a similarlity to the case of quantum S$^2$. It is known that the 
$q \rightarrow 1$ limit of quantum S$^2$ is the ordinary fuzzy 
S$^2$.\cite{qSphere}\cite{qsphereWZ}
To discuss $q \rightarrow 1$ limit in the present case it is better to 
rescale the matrices as $(X, Y, Z) \rightarrow (\hbar X', 
\hbar Y', \hbar Z')$. Then the relations (\ref{FT1})-(\ref{FT3}) take the 
form.  
\begin{eqnarray}
\ [ X', Y'] &=& i  Z', \nonumber  \\
\ [ Y', Z'] & = & i \left\{X', \ \hbar^2({X'}^2+{Y'}^2) -\mu \right\}, 
\nonumber  \\
\ [ Z', X'] &=& i \left\{Y', \ \hbar^2({X'}^2+{Y'}^2) -\mu \right\}, 
\label{rescaledFT} 
\end{eqnarray}
In the $\hbar \rightarrow 0$ limit  
this reduces to SU(2) or SU(1,1) algebra 
according to the sign of $\mu$\footnote{SU(2) for $\mu <0$ and SU(1,1) for 
$\mu >0$ after trivial rescaling of $X'$, $Y'$, $Z'$. }
Therefore when $\mu <0$, the classical limit $\hbar \rightarrow 0$ can be 
taken and fuzzy S$^2$ is obtained. When $\mu >0$, the limiting algebra does 
not describe fuzzy S$^2$ nor T$^2$ but hyperboloid. 

Under the above rescaling the q-deformed Lie algebra takes the form.
\begin{eqnarray}
\ [ \ D'-\frac{1}{q}\tilde{D}', \  W' \, ]_q &=& 
-\frac{(q+1)^2\mu}{q} \ W', \nonumber \\
\ [ \ D'-q\tilde{D}', \  W'^{\dagger} \, ]_q &=& 
(q+1)^2\mu \ {W'}^{\dagger},
\nonumber \\
 \ [ \ W', \ D'-q\tilde{D}' \, ]_q &=& 
(q+1)^2\mu \ W' , \nonumber \\
\ [ \ {W'}^{\dagger}, \ D'-\frac{1}{q} \tilde{D}' \, ]_q &=& 
-\frac{(q+1)^2\mu}{q} \ {W'}^{\dagger},
\nonumber \\
\ [W', \ {W'}^{\dagger}]_q &=& qD'-\tilde{D}', \nonumber \\
\ [{W'}^{\dagger}, \ W']_q &=& q \tilde{D}'-D', \nonumber \\
\ [D', \tilde{D}'] &=& 0,  \nonumber \\
D'+\tilde{D}' &=&W'{W'}^{\dagger}+{W'}^{\dagger}W', \nonumber \\
c \bm{1} 
&=& \frac{1}{4}(D'+\tilde{D}'-2\mu)^2+\frac{1}{4\hbar^2}(D'-\tilde{D}')^2
\label{rescalednewalg}
\end{eqnarray}
Here we set $(W, W^{\dagger}, D, \tilde{D}) \rightarrow 
(\hbar W', \hbar {W'}^{\dagger}, \hbar D', \hbar \tilde{D}')$. 
If we let $q \rightarrow 1$, the algebra becomes
\begin{eqnarray}
\ [D'-\tilde{D}', \ W'] &=& -4\mu W', \nonumber \\
\ [D'-\tilde{D}', \ {W'}^{\dagger}] &=& 4\mu {W'}^{\dagger}, \nonumber \\
\ [W', {W'}^{\dagger}] &=& D'-\tilde{D}', \nonumber \\
\ [D', \tilde{D}']=0
\label{su2}
\end{eqnarray}
In addition, the ${\cal O}(q-1)$ terms of the first eq of 
(\ref{rescalednewalg}) yield 
\begin{equation}
D'W'-W'\tilde{D}'=0.
\end{equation}
This last equation determines the matrix $D'+\tilde{D}'$ decoupled from 
(\ref{su2}). The algebra (\ref{su2}) except for the last eq is an SU(2) 
(or SU(1,1)) Lie algebra according to the sign of $\mu$. 
Therefore for $q \neq 1$ and  $\mu <0$ the algebra 
(\ref{rescalednewalg}) will describe a q-deformation 
of the ordinary fuzzy sphere. We note that the algebra 
(\ref{rescalednewalg}) also makes sense for real values of $q$.
This case, however, will not be considered in this paper. 

Now the N-dimensional irreducible, unitary representation of the algebra 
(\ref{rescalednewalg}) will be presented. 
This corresponds to 'string solution' in \cite{FRS} and exists for generic values of $q$. 
Although the representations of 
(\ref{FT1})-(\ref{FT4}) were obtained in \cite{FRS}, some relations among 
$q$ ($\hbar$), $ \ N$, $\mu$ were assumed there. This is because the 
invariant $c$ was set to 1 in (\ref{FT4}). In this paper we will take 
$c$ as a free parameter to be determined later and $q$ will be treated as 
independent. 

We will take $D'$, $\tilde{D}'$ to be diagonal. As in the case of classical 
SU(2) algebra the matrix elements of $W'$, ${W'}^{\dagger}$ are assumed to be 
\begin{eqnarray}
W'_{mn} &=& a_m \ \delta_{m+1.n}, \nonumber \\
(W'^{\dagger})_{mn} &=& a_{m-1} \ \delta_{m,n+1}. \quad (m,n=1,2,..,N)
\label{29}
\end{eqnarray}
Here we set $a_0=a_N=0$. 
By (\ref{DD}) the diagonal components of $D'$, $\tilde{D}'$ are given by
\begin{equation}
D'_n=|a_n|^2, \qquad \tilde{D}'_n= |a_{n-1}|^2. 
\end{equation}
Then the first relations in (\ref{rescalednewalg}) 
yield the following equation.
\begin{equation}
|a_{n+1}|^2-(q+q^{-1}) \ |a_n|^2+|a_{n-1}|^2=\frac{(q+1)^2}{q} \ \mu
\label{31}
\end{equation}
This equation is solved with the initial condition $a_0=0$.
\begin{equation}
|a_n|^2=\frac{q(q^n-q^{-n})}{q^2-1} \ |a_1|^2+ \frac{(q+1)(q^{n}+q^{-n+1}-q
-1)}{(q-1)^2} \ \mu
\end{equation}
Then the condition $a_N=0$ determines $a_1$. 
When $\mu \neq 0$, the result for $a_n$ is 
\begin{equation}
|a_n|^2= \frac{(q+1)^2 (q^n-q^N)(q^n-1)}{(q-1)^2q^n(q^N+1)} \ \mu. 
\label{an}
\end{equation}
The case $\mu=0$ requires care and the solution exists only if $q^{2N}=1$,  
and is given by\cite{FRS} 
\begin{equation}
|a_n|^2=\frac{q^n-q^{-n}}{q-q^{-1}} \ |a_1|^2. 
\end{equation}

Note that in the above solution (\ref{an}) no relation among $q$, N and 
$\mu$ is assumed. The condition $|a_n|^2 \geq 0$, however, restricts the 
allowed values of N for fixed $q$. For given N this condition in turn 
restricts the range of $q+q^{-1}$.  Let us present some examples. \\
When $N=2$, 
\begin{equation}
|a_1|^2=-\frac{q+q^{-1}+2}{q+q^{-1}} \ \mu.
\end{equation}
Because $-2 \leq q+q^{-1} \leq 2$ for $|q|=1$, the allowed value of $q+q^{-1}$ 
is $-2 \leq q+q^{-1} <0$ for $\mu >0$, and $0 < q+q^{-1} \leq 2$ 
for $\mu <0$. \\
When $N=3$, 
\begin{equation}
|a_1|^2=|a_2|^2= -\frac{q+q^{-1}+2}{q+q^{-1}-1} \ \mu.
\end{equation}
In this case $-2 \leq q+q^{-1} <1$ for $\mu >0$, and $1 < q+q^{-1} \leq 2$ 
for $\mu <0$. \\
When $N=4$, 
\begin{eqnarray}
|a_1|^2=|a_3|^2 &=& -\frac{(q+q^{-1}+2)(q+q^{-1}+1)}{(q+q^{-1})^2-2} \ \mu, 
\nonumber \\
|a_2|^2=|a_4|^2 &=& -\frac{(q+q^{-1}+2)^2}{(q+q^{-1})^2-2} \ \mu. 
\end{eqnarray}
So $-1 \leq q+q^{-1} < \sqrt{2}$ for $\mu >0$ and $\sqrt{2} < q+q^{-1} \leq 2$
for $\mu <0$. 
In all the above cases the $q \rightarrow 1$ limit is allowed only for 
$\mu <0$. This is in accord with the above comment on this limit. 
When $\mu=0$, the solution which satisfy $|a_n|^2 >0$ is 
$q=\exp (\pm \frac{\pi i}{N})$ and $|a_n|^2= |a_1|^2 \ \sin \frac{n\pi}{N}/ 
\sin \frac{\pi}{N}$. 

Finally, the invariant $c$ (\ref{CMetric}) (for $\mu \neq 0$) is  obtained as
\begin{eqnarray}
c &=&\frac{\hbar^4}{4} \ 
\left(|a_n|^2+|a_{n-1}|^2-\frac{2\mu}{\hbar^2} \right)^2+ \frac{\hbar^2}{4} \ 
\left(|a_n|^2-|a_{n-1}|^2 \right)^2 \nonumber \\ 
&=& \mu^2 \ \frac{q+q^{-1}+2}{q^N+q^{-N}+2}.
\label{c}
\end{eqnarray}
The final result does not depend on $n$, as it should. 

As mentioned at the beginning of this paper fuzzy T$^2$ is realized for 
$\mu >0$ and $c < \mu^2$. This last equation yields a condition on the range 
of $q+q^{-1}$. 
For example, when $N=2$, the condition $q+q^{-1} < q^2+q^{-2}$ is satisfied 
for $q+q^{-1}<-1$.
Combining this with the condition for $\mu>0$ obtained above, we obtain the 
allowed range for fuzzy T$^2$, $-2 \leq q+q^{-1} <-1$. 
When $N=3$, the allowed range for fuzzy T$^2$ is $-2 \leq q+q^{-1} <0$. 
When $N=4$, it is $-1 \leq q+q^{-1} < (\sqrt{5}-1)/2$.  

In the classical limit $q \rightarrow 1$ the invariant $c$ (\ref{c}) 
approaches $\mu^2$. The condition (\ref{Casimir}) on the rescaled matrices 
$X'$, $Y'$, $Z'$ becomes in this limit 
\begin{equation}
-2\mu \left(X'^2+Y'^2\right)+Z'^2= \lim_{\hbar \rightarrow 0} 
\frac{c-\mu^2}{\hbar^2}= \mu^2 \ (N^2-1). 
\end{equation}
When we set $N=2j+1$, the right-hand side becomes $j(j+1) \ (2\mu)^2$. 
For $\mu <0$ after trivial rescaling of $X'$, $Y'$, $Z'$, this coincides 
with the condition for fuzzy S$^2$ in the spin $j$ representation.

To summarize it has been shown that the fuzzy T$^2$ algebra presented in 
\cite{FRS} can be rearranged into a different algebra 
(\ref{rescalednewalg}), which takes the form of a q-deformed Lie algebra. 
This may make it possible to treat fuzzy T$^2$ by arguments based on symmetry
principles.  There is a barrier at $\mu=0$. The algebra for $\mu <0$ is 
SU(2)-like but that for $\mu >0$ is SU(1,1)-like. 
The classical $q \rightarrow 1$ limit of the squashed S$^2$ with $\mu <0$ is 
fuzzy S$^2$, but the classical limit of the fuzzy torus which belongs to 
$\mu > 0$ does not exist. For $q$ not a root of unity 
irreducible N-dimensional representations (``string representation'') 
of the q-deformed Lie algebra (\ref{rescalednewalg}) are obtained and 
the allowed range for $q+q^{-1}$ which 
depends on the sign of $\mu$ is obtained for some values of N. The allowed 
range for $q+q^{-1}$ which describes fuzzy T$^2$ is also obtained. 
It is not clear whether a quantum group associated with the deformed 
Lie algbera (\ref{rescalednewalg}) exists. If it does, for further 
investigation and construction of field theories on fuzzy T$^2$ 
it will be necessary to study the representations of the quantum group.  

\section*{Note added}
In addition to the representation (\ref{29}), (\ref{an}) there also exits for 
q a root of unity ($q^N=1$) a solution called ``loop solution'' \cite{FRS} with 
\begin{equation}
W'_{mn}=a_m \delta_{m+1,n} \ \ \  (mod \ \ N).
\label{annew}
\end{equation}
$a_m$ is periodic; $a_{m+N}=a_m$. In \cite{FRS} it was discussed that this 
'loop solution' may be important. In this note this representation for 
$q^N=1$ will be briefly discussed.
 
As will be shown soon this solution is not uniquely determined by the algebra 
(\ref{FT1})- (\ref{FT4}) or (\ref{rescalednewalg}). It contains two free 
parameters. So this solution does not determine  fuzzy T$^2$ or S$^2$ 
completely. This may bring about a problem when the matrix model for these 
surfaces is formulated. The number of the classical solutions to the eqs of 
motion in the matrix model will be continuously infinite with the solutions 
depending on the free parameters. So throughout the main text of this paper 
we discussed only the representations with generic values of $q$. 

With the ansatz (\ref{annew}) for $W'$ the recursion relation for $a_n$ is 
still given by (\ref{31}). Defining $b_n$ by 
\begin{equation}
b_n = |a_n|^2+ \frac{(q+1)^2}{(q-1)^2} \ \mu
\label{41}
\end{equation}
and solving the relation $b_{n+1}-qb_n=q^{-1}(b_n-qb_{n-1})$, we obtain
\begin{equation}
b_n=\frac{q^{1-n}-q^{1+n}}{1-q^2} \ b_1+ \frac{q^n-q^{2-n}}{1-q^2} \ b_0.
\label{42}
\end{equation}
It will be assumed that $q^2 \neq 1$. 
The periodicity conditions $b_N=b_0$, $b_{N+1}=b_1$ lead to 
\begin{eqnarray}
(q^N-q^{2-N}+q^2-1) \ b_0+(q^{1-N}-q^{1+N}) \ b_1 &=& 0,  \nonumber \\
(q^{N+1}-q^{1-N}) \ b_0+(q^{-N}-q^{N+2}+q^2-1) \ b_1 &=& 0 
\end{eqnarray}
The determinant of the coefficient matrix vanishes only for $q^N=1$. 
When $q^N \neq 1$, these eqs have only the solution $b_0=b_1=0$ and by 
(\ref{42}) all $b_n=0$. This gives a collasped ($Z=0$) surface.\cite{FRS} 
Instead when $q^N=1$, $b_n$ and hence $a_n$ is expressed in terms of $a_0$, 
$a_1$ by (\ref{41}), (\ref{42}). 
\begin{eqnarray}
|a_n|^2 &=& \frac{q^{1-n}-q^{1+n}}{1-q^2} \ |a_1|^2+ \frac{q^n-q^{2-n}}{1-q^2} \ |a_0|^2 
\label{44}
\nonumber \\
&& + \mu \frac{q+1}{(q-1)^2} \ (q^n+q^{1-n}-1-q)
\end{eqnarray}
The two parameters $a_0$, $a_1$ are left undetermined. The ``loop solution'' 
of \cite{FRS} contains a free parameter $\beta$. (eq(5.22) of this reference)

We will remark on the ``boundary'' between representations for torus and 
sphere. 
In this paper the invariant $c$ (\ref{CMetric}) is not set to 1 but treated as a parameter to be 
determined by the representation matrices. 
The invariant $c$ (\ref{CMetric}) for 
this ``loop solution'' (\ref{44}) is given by 
\begin{eqnarray}
c &=& -\frac{q(1-q)^2}{(1+q)^4} \ \left\{ |a_0|^4+|a_1|^4-(q+q^{-1})|a_0|^2 |a_1|^2 \right. 
\nonumber \\
&& \left. -\mu q^{-1}(q+1)^2 (|a_0|^2+|a_1|^2) \right\} + \mu^2
\end{eqnarray}
For $c > \mu^2$ and $\mu >0$ one obtains fuzzy T$^2$ and for $c< \mu^2$ 
fuzzy S$^2$.  
The boundary is determined by $c-\mu^2=0$ 
and is a {\em quadratic curve} on the ($|a_0|^2, \ |a_1|^2$) plane. There are some 
special points on this curve. 
At $a_0=a_1=0$ the reduction of the dimensionality of the representation 
($N \rightarrow N-1$) takes place \cite{FRS} and this was argued to be a 
singularity. One can show there also exist points that correspond to 
$a_M=a_{M+1}=0$, ($M=1,2, ..., N-1$) on this curve. 
Yet other points on the curve are nonsingular. 

\section*{Acknowledgments}
\hspace{5mm}
The author thanks H.~Fuji for discussion. 
This work is supported in part by Grant-in-Aid (No.13135201) 
from the Ministry of Education, Science, Sports and Culture of Japan 
(Priority Area of Research (2)).


\newpage

\begin{thebibliography}{99}
\bibitem{Madore} J.~Madore, Class. Quantum Geom. {\bf 9} (1992) 69. 
\bibitem{JH} J.~Hoppe, {\it Quantum Theory of a Relativistic Surface}, 
Ph. D. Thesis;J.~Hoppe, {\it Diffeomorphism groups, quantization and 
SU($\infty$)}, Int.J.Mod. Phys. {\bf A4} (1989) 5235.
\bibitem{gaugeaction} A.~Y.~Alekseev, A.~Recknagel and V.~Schomerus, {\it Brane Dynamics in Background Fluxes and Non-commutative Geometry}, hep-th/0003187.
\bibitem{Presnejder} P.~Pre\v{s}najder, {\it The origin of chiral anomaly and
the noncommutative geometry}, hep-th/9912050.
\bibitem{HNT} K.~Hayasaka, R.~Nakayama and Y.~Takaya, {\it A new 
noncommutative product on the fuzzy two-sphere corresponding to the unitary 
representation of $SU(2)$ and the Seiberg-Witten map}, Phys. Lett. {\bf B553} 
(2003)109-118, hep-th/0209240.
\bibitem{CP2} 
V.~P.~Nair and S.~Randjbar-Daemi, {\it On brane solutions in M(atrix) theory},
 Nucl. Phys. {\bf B533} (1998) 333-347, hep-th/9802187; \\
H.~Grosse and A.~Strohmaier, {\it Towards a nonperturbative covariant 
reguralization in 4d field thery}, Lett. Math. Phys. {\bf 48} (1999) 
163-179, hep-th/9902138; \\
G.~Alexanian, A.~P.~Balachandran, G.~Immirzi and B.~Ydri, {\it Fuzzy CP$^2$},
J. Geom. Phys. {\bf 42} (2002) 28-53; \\
A.~Balachandran, B.~Dolan, J.~Lee, X.~Martin and D.~O'Conner, {\it Fuzzy 
complex projective spaces and their star-products}, J. Geom. Phys. {\bf 43} 
(2002) 184-204, hep-th/0107099; \\
T.~Azuma, S.~Bal, K.~Nagao and J.~Nishimura, {\it Dynamical aspects of the 
fuzzy CP$^2$ in the large N reduced model with a cubic term}, 
hep-th/0405277; \\
H.~Grosse and H.~Steinacker, {\it Finite gauge theory on fuzzy CP$^2$}, 
hep-th/0407089. 
\bibitem{FT} A.~Conne, M.~R.~Douglas and A.~Schwarz, {\it Noncommutative Geometry and Matrix Theory: Compactification on Tori}, hep-th/9711162; \\
J.~Ambjorn, Y.~M.~Makeenko, J.~Nishimura and R.~J.~Szabo, {\it Lattice Gauge Fields and Discrete Noncommutative Yang-Mills Theory}, hep-th/0004147; \\
S.~Bal, M.~Hanada, H.~Kawai and F.~Kubo, {\it Fuzzy Torus in Matrix Model}, Nucl.Phys. {\bf B727} (2005) 196, hep-th/0412303; \\
D.~Biagatti, {\it String Thoery and the Fuzzy Torus}, Int.J.Mod.Phys. {\bf A19} (2004) 4533-4566, hep-th/0405115.
\bibitem{FRS} J.~Arnlind, M.~Bordemann, L.~Hofer, J.~Hoppe and H.~Shimada,
{\it Fuzzy Riemann Surfaces}, hep-th/0602290. 
\bibitem{qSphere} P.~Podle\'{s}, Lett. Math. Phys. {\bf 14} (1987) 193; \\
H.~Grosse, J.~Madore and H.~Steinacker, {\it Field Theory on the q-deformed Fuzzy Sphere I}, hep-th/0005273; \\ 
H.~Steinacker, {\it Quantum Field Theory on the q-deformed Fuzzy Sphere}, hep-th/0105126.
\bibitem{qsphereWZ} A.~Y.~Alekseev, A.~Recknagel and V.~Shomerus, {\it Non-commutative World-volume Geometries: Branes on SU(2) and Fuzzy Spheres}, hep-th/9908040.
\end{thebibliography}
\end{document}